\documentclass{elsart}
\usepackage{amssymb}
\usepackage{amsfonts}
\usepackage{amsmath}

\input{tcilatex}

\begin{document}

\begin{frontmatter}%

\title{Exact renormalization group equation for the Lifshitz critical point}
%

\author{C.\ Bervillier}%

\ead{bervil@spht.saclay.cea.fr}
%

\address{Service de physique th\'{e}orique,
CEA/DSM/SPhT-CNRS/SPM/URA 2306
CEA/Saclay,
F-91191 Gif-sur-Yvette C\'{e}dex, France}
%

\begin{abstract}
An exact renormalization equation (ERGE) accounting for an anisotropic
scaling is derived. The critical and tricritical Lifshitz points are then
studied at leading order of the derivative expansion which is shown to
involve two differential equations. The resulting estimates of the Lifshitz
critical exponents compare well with the $O\left( \epsilon ^{2}\right) $
calculations. In the case of the Lifshitz tricritical point, it is shown
that a marginally relevant coupling defies the perturbative approach since
it actually makes the fixed point referred to in the previous perturbative
calculations $O\left( \epsilon \right) $ finally unstable.%
\end{abstract}%


\end{frontmatter}%

An exact renormalization group (RG) equation (ERGE) is a continuous version
of the (originally discrete) Wilson RG transformation of an action $S\left[
\phi \right] $ \cite{440}. It results in an integro-differential equation
for the action on which one may apply nonperturbative approximations or
truncations such as an expansion in powers of the derivative of the field $%
\phi $ (for an introductory review see \cite{4595}). In a preceding paper 
\cite{5744} I have argued that the study of the first terms of the
derivative expansion of an ERGE may favorably compete with the customary
perturbative approach provided that one does not expect an accuracy larger
than that obtained from the $\epsilon $-expansion developed up to, say, $%
O\left( \epsilon ^{2}\right) $ and even, sometimes, one could show that the
perturbative approach has failed. The Lifshitz critical point \cite{5247}
(for reviews see \cite{5629,4764,5296}) is an excellent example to test
these claims. The previous calculations relative to this system have been
done up to $O\left( \epsilon ^{2}\right) $ only. They have shown the
difficulty of estimating some Feynman graph integrals, various results have
coexisted for a long time and a controversy still persists (see the recent
review in \cite{5296}). Moreover, the tricritical side involves several
operators among which one is marginal at the physical dimension and has not
been (cannot be) included in the previous perturbative studies \cite%
{5683,5680}.

The object of the present study is threefold. In a first time I write down
the ERGE adapted to the study of the Lifshitz point. The main difficulty in
that case is the existence of two distinct correlation lengths which induce
an anisotropic scaling. In a second time I derive the leading order of the
derivative expansion applied to the ERGE. In a third time I numerically
study the differential equations that follow using the shooting method as in 
\cite{5744}.

\section{Derivation of the ERGE}

\label{section I}

A Lifshitz critical point is a multicritical point at which a disordered, a
homogeneous ordered and a (spatially) modulated ordered phases meet \cite%
{5247}. The simplest way of illustrating the theoretical conditions of
realization of a Lifshitz point is to look at the Landau form of the
Euclidean action (Hamiltonian) relevant to the problem: 
\begin{equation}
S\left[ \phi \right] _{\mathrm{Landau}}=\int \mathrm{d}^{d}x\left[ \left(
\partial _{\perp }\phi \right) ^{2}+\rho \left( \partial _{\parallel }\phi
\right) ^{2}+\tau \phi ^{2}+u\phi ^{4}+\sigma \left( \partial _{\parallel
}^{2}\phi \right) ^{2}\right]  \label{eq:Landau}
\end{equation}
in which $\phi \left( x\right) $ is a scalar field (the generalization to a $%
n$-vector field is straightforward), $\partial _{\perp }$ and $\partial
_{\parallel }$ stand for the derivatives along two directions ($x_{\parallel
}$ and $x_{\perp }$) in the $x$-space of dimension $d$. The subspace $%
x_{\parallel }$ involves $m$ components while the remaining $x_{\perp }$ has 
$\left( d-m\right) $ components. The Lifshitz critical point corresponds to
both $\tau =0$ and $\rho =0$ (in the Landau approximation, in general $\tau
=\tau _{0}$ and $\rho =\rho _{0}$). Usually one would have named this point
a tricritical point however the qualifier tricritical associated to Lifshitz
has been reserved to the case where $u$ vanishes also with $\tau $ and $\rho 
$ ---in that case other terms must be added \cite{5680,5685} to (\ref%
{eq:Landau}), such as $c_{1}\int \mathrm{d}^{d}x\phi ^{6}$ and $c_{2}\int 
\mathrm{d}^{d}x\left( \partial _{\parallel }\phi \right) ^{2}\phi ^{2}$.

Lifshitz criticality is characterized by an anisotropic scaling. The
expected scaling property of the two-point vertex function (in the momentum
space) is as follows \cite{5296} ($s\rightarrow 0$): 
\begin{eqnarray}
\Gamma ^{\left( 2\right) }\left( sq_{\perp },0\right) &\sim &s^{2-\eta _{L2}}
\\
\Gamma ^{\left( 2\right) }\left( 0,sq_{\parallel }\right) &\sim &s^{4-\eta
_{L4}} \\
\Gamma ^{\left( 2\right) }\left( sq_{\perp },s^{\theta }q_{\parallel
}\right) &\sim &s^{2-\eta _{L2}}  \label{eq:ScalingBehav}
\end{eqnarray}
in which $\eta _{L2}$ and $\eta _{L4}$ vanish in the Landau approximation
and $\theta =\frac{ 2-\eta _{L2}}{ 4-\eta _{L4}} $. The general expected
scaling property (at the fixed point) is \cite{4767,5629}: 
\begin{equation}
\Gamma ^{\left( 2\right) }\left( s^{\lambda _{\tau }}\mu _{\tau },s^{\lambda
\rho }\mu _{\rho };sq_{\perp },s^{\theta }q_{\parallel }\right) =s^{2-\eta
_{L2}}\Gamma ^{\left( 2\right) }\left( \mu _{t},\mu _{\rho };q_{\perp
},q_{\parallel }\right)  \label{eq:generalscaling}
\end{equation}
in which $\mu _{t}$ and $\mu _{\rho }$ are the two relevant parameters
(scaling fields) associated to $\tau $ and $\rho $ respectively, $\lambda
_{\tau }=1/\nu _{L2}$ and $\lambda _{\rho }=\varphi /\nu _{L2}$ where $\nu
_{L2}$ is the critical exponent of the ``perpendicular'' correlation length (%
$\xi _{\perp } $) and $\varphi $ the crossover exponent controlling the
crossing between the Lifshitz and ordinary critical points.

The adaptation of the Wilson-Polchinski ERGE to the problem is in fact an
easy task: it is a straightforward application of the rules stated in \cite%
{5744} to the case involving two sets of scales ($\parallel $ and $\perp $).

Let me consider the following general action with a double cutoff function $%
P $: 
\begin{equation}
S\left[ \phi \right] =\frac{1}{2}\int_{q}\phi _{q}P^{-1}(\frac{q_{\perp }^{2}%
}{\Lambda _{\perp }^{2}},\frac{q_{\shortparallel }^{2}}{\Lambda _{\parallel
}^{2}};\ell _{\perp },\ell _{\parallel })\phi _{-q}+S_{\mathrm{int}}\left[
\phi \right]
\end{equation}
with $\vec{q}_{\parallel }=q_{1},\cdots ,q_{m}$, $\vec{q}_{\perp
}=q_{m+1},\cdots ,q_{d}$, $m\in \left[ 0,d\right] $. The two running cutoffs 
$\Lambda _{\perp }$ and $\Lambda _{\parallel }$ are related to two
independent momentum scales of reference $\Lambda _{0\perp }$ and $\Lambda
_{0\parallel }$ via: $\Lambda _{\perp }=\ell _{\perp }\Lambda _{0\perp }$
and $\Lambda _{\parallel }=\ell _{\parallel }\Lambda _{0\parallel }$ with
the anisotropy relation: 
\begin{equation}
\ell _{\parallel }=\ell _{\perp }^{\theta }  \label{eq:anisorel}
\end{equation}
I underline the fact explained in \cite{5744} that, contrary to the current
use, the cutoff function must not depend only on the ratio $q^{2}/\Lambda
^{2}$ but also explicitly on $\ell $ (here on $\ell _{\parallel }$ and $\ell
_{\perp }$) in order to correctly account for the ``history'' of the many
scale effect which little by little starting from some large momentum scale $%
\Lambda _{0}$ (here $\Lambda _{0\parallel }$ and $\Lambda _{0\perp }$)
finally induces the anomalous scaling observed at small momenta [here given
by eq. (\ref{eq:ScalingBehav})].

In the following, for the sake of simplifying the notations, I shall use $%
\ell _{\perp }=\ell =\mathrm{e}^{-t}$.

Due to the anisotropy relation (\ref{eq:anisorel}) the explicit double
dependence of $P$ on $\ell _{\perp }$ and $\ell _{\parallel }$ reduces to a
dependence on the unique renormalization parameter $\ell $, so that in the
following I shall write $P(\frac{q_{\perp }^{2}}{\Lambda _{\perp }^{2}},%
\frac{q_{\shortparallel }^{2}}{\Lambda _{\parallel }^{2}};\ell )$.

As explained in \cite{5744} and repeated above, the explicit $\ell $%
-dependence in $P$ is essential. Technically speaking, it is required to get
a fixed point relevant to the physics under study while the resulting ERGE
does not depend explicitly on $\ell $. Specifically, the ``physics'' which
one is presently interested in is merely the expected scaling behavior (\ref%
{eq:ScalingBehav}) and thus, following the same considerations as in \cite%
{5744} one easily shows that the $\ell $-dependence of $P$ must factorize: 
\begin{equation}
P(\frac{q_{\perp }^{2}}{\Lambda _{\perp }^{2}},\frac{q_{\shortparallel }^{2}%
}{\Lambda _{\parallel }^{2}};\ell )=\ell ^{\varpi }\tilde{P}(\frac{q_{\perp
}^{2}}{\Lambda _{\perp }^{2}},\frac{q_{\shortparallel }^{2}}{\Lambda
_{\parallel }^{2}})
\end{equation}
with 
\begin{equation}
\varpi =1-\eta _{L2}/2
\end{equation}

Then the ERGE adapted to the anisotropic scaling of interest follows ($\dot{S%
}=dS/dt=-\ell dS/d\ell =-\theta \ell _{\parallel }dS/d\ell _{\parallel }$): 
\begin{eqnarray}
\dot{S} &=&{\mathcal{G}}_{\mathrm{dil}}S  \notag \\
&&-\int_{q}\left( \varpi \tilde{P}-q_{\perp }^{2}\tilde{P}_{\perp }^{\prime
}-\theta q_{\parallel }^{2}\tilde{P}_{\parallel }^{\prime }\right) \left[ 
\frac{\delta S}{\delta \phi _{q}}\frac{\delta S}{\delta \phi _{-q}}-\frac{%
\delta ^{2}S}{\delta \phi _{q}\delta \phi _{-q}}-2\tilde{P}^{-1}\phi _{q}%
\frac{\delta S}{\delta \phi _{q}}\right]  \label{eq:ERGELifshitz}
\end{eqnarray}
in which $q$, $q_{\perp }$, $q_{\parallel }$ are dimensionless and $\tilde{P}%
_{\perp ,\parallel }^{\prime }=\mathrm{d}\tilde{P}\left( q_{\perp
},q_{\parallel }\right) /\mathrm{d}q_{\perp ,\parallel }$ and: 
\begin{eqnarray}
{\mathcal{G}}_{\mathrm{dil}}S &=&-\int_{q}\phi _{q}\,\left( \frac{d_{m}}{2}%
+\varpi +\mathbf{q}_{\perp }\cdot \partial _{q_{\perp }}+\theta \mathbf{q}%
_{\parallel }\cdot \partial _{q\parallel }\right) \frac{\delta S}{\delta
\phi _{q}}  \label{eq:Gdil} \\
d_{m} &=&d+m\left( \theta -1\right)  \label{eq:dm}
\end{eqnarray}

Notice that in the case of the ordinary critical point ($m=0$) the above
equations give back the ordinary Wilson-Polchinski ERGE of \cite{5744} while
in the isotropic Lifshitz case ($m=d$) the renormalization parameter to be
considered is $\ell _{\parallel }$ instead of presently $\ell =\ell _{\perp
} $ and one must account for (\ref{eq:anisorel}) to get the ERGE adapted to
the isotropic Lifshitz point from the eqs (\ref{eq:ERGELifshitz}-\ref{eq:dm}%
) [practical rules: set formally $\theta =1$ and change $\varpi \rightarrow 
\frac{\varpi }{\theta }$].

\section{Leading order of the derivative expansion}

The object of this section is to obtain the leading order of the derivative
expansion of eqs (\ref{eq:ERGELifshitz}-\ref{eq:dm}). To this end it is
worth reminding the reader with the usual case of the ordinary critical
point ($m=0$). In that case, the leading order of the derivative expansion
is named the local potential approximation and is associated with the
critical exponent $\eta =0$. One knows that it consists of a truncation of
the ERGE to actions of the type: 
\begin{equation}
S_{\mathrm{LPA}}=\int \mathrm{d}^{d}x\left[ Z\left( \partial \phi \right)
^{2}+V\left( \phi \right) \right]
\end{equation}
with $Z$ a constant independent of $\phi $ and of $t$ ($=-\ln \ell $). Eq (%
\ref{eq:ERGELifshitz}) then reduces to a single equation for the unique
function $V$ \cite{note1}: 
\begin{equation}
\dot{V}=dV-\left( \frac{d}{2}-\varpi \right) \phi V^{\prime }-\varpi \left(
V^{\prime 2}-V^{\prime \prime }\right)  \label{eq:LPAOrdin}
\end{equation}
and the fact that $Z$ is constant (i.e. $\dot{Z}=0$) implies the condition $%
\varpi =1$ (i. e. $\eta =0$). The next order in the derivative expansion is
then obtained by allowing $Z$ to depend on $\phi $ and $\eta $ no longer
vanishes.

By analogy, the lowest approximation in the case of the isotropic Lifshitz
point ($m=d$), leads to assign the role played by $Z$ to the coefficient $A$
of a term $\left( \partial ^{2}\phi \right) ^{2}$ in the action $S$ while $Z$
is no longer constrained: 
\begin{equation}
S_{\mathrm{Liso}}=\int \mathrm{d}^{d}x\left[ Z\left( \phi \right) \left(
\partial \phi \right) ^{2}+V\left( \phi \right) +A\left( \partial ^{2}\phi
\right) ^{2}\right]
\end{equation}
There are two equations ($\dot{Z}$ and $\dot{V}$) and the condition $A=%
\mathrm{const}$ implies that $\eta _{L4}=0$.

In the case of an $m$-axial Lifshitz point, the leading order of the
derivative expansion is a mixing of the two previous cases such that the
approximation corresponds to both $\eta _{L2}=0$ and $\eta _{L4}=0$. Hence,
the action is limited to the following form: 
\begin{equation}
S_{\mathrm{Lm}}=\int \mathrm{d}^{d}x\left[ Z_{\perp }\left( \partial _{\perp
}\phi \right) ^{2}+Z_{\parallel }\left( \phi \right) \left( \partial
_{\parallel }\phi \right) ^{2}+V\left( \phi \right) +A\left( \partial
_{\parallel }^{2}\phi \right) ^{2}\right]
\end{equation}
yielding two equations for $Z_{\parallel }\left( \phi \right) $ and $V\left(
\phi \right) $ while $Z_{\perp }$ and $A$ are constants independent of $\phi 
$ and $t$ what implies $\varpi =1$ and $\theta =1/2$ (i. e., $\eta
_{L2}=\eta _{L4}=0$).

Following \cite{5744}, before writing down these equations let me introduce
some convenient modifications. First, in the complete equation, I consider a
redundant transformation of the field $\tilde{\phi}_{q}=\psi \left(
q^{2}\right) \phi _{q}$ in which $\psi \left( q^{2}\right) $ is arbitrary
except the normalization $\psi \left( 0\right) =1$, then I subtract the high
temperature fixed point $\frac{1}{2}\int_{p}\left( \tilde{P}\psi ^{2}\right)
^{-1}\phi _{p}\phi _{-p}$ from the action. Two other useful modifications
are introduced in the derivative expansion. First I rescale the field $\phi
=I_{0}^{1/2}\bar{\phi}$ and the potential $V=I_{0}\bar{V}$ where $%
I_{0}=\int_{q}\left( \varpi \tilde{P}-q_{\perp }^{2}\tilde{P}_{\perp
}^{\prime }-q_{\parallel }^{2}\tilde{P}_{\parallel }^{\prime }\right) \psi
^{2}$ and instead of $V$, I consider the equation for $\mathrm{v}_{1}=%
\mathrm{d}\bar{V}/\mathrm{d}\bar{\phi}$. Finally, for the sake of unified
notations I rename $Z_{\parallel }$ as $\mathrm{v}_{2}$. With these new
definitions, the RG equations in the leading order of the derivative
expansion read: 
\begin{eqnarray}
\mathrm{\dot{v}}_{1} &=&\mathrm{v}_{1}^{\prime \prime }{+}d_{\varphi }%
\mathrm{v}_{1}-\left( \tilde{d}_{\varphi }\bar{\phi}+2\varpi \mathrm{v}%
_{1}\right) \mathrm{v}_{1}^{\prime }+P_{1}\mathrm{v}_{2}^{\prime }
\label{eq:1} \\
\mathrm{\dot{v}}_{2} &=&\mathrm{v}_{2}^{\prime \prime }-2\left( \varpi
+\theta \right) \mathrm{v}_{2}-\left( \tilde{d}_{\varphi }\bar{\phi}+2\varpi 
\mathrm{v}_{1}\right) \mathrm{v}_{2}^{\prime }  \notag \\
&&+\mathrm{v}_{1}^{\prime }\left( P_{2}\mathrm{v}_{1}^{\prime }-2\theta \psi
_{\parallel 0}^{\prime }-4\varpi \mathrm{v}_{2}\right)   \label{eq:2}
\end{eqnarray}%
in which the prime means the derivative with respect to $\bar{\phi}$, $%
d_{\varphi }=\frac{d_{m}}{2}-\varpi $, $\tilde{d}_{\varphi }=\frac{d_{m}}{2}%
+\varpi $, $\psi _{\parallel 0}^{\prime }=\left. \mathrm{d}\psi /\mathrm{d}%
q_{\parallel }^{2}\right\vert _{q=0}$, $P_{1}=2\frac{I_{1}}{I_{0}}$ with $%
I_{1}=\int_{q}q^{2}\left( \varpi \tilde{P}-q_{\perp }^{2}\tilde{P}_{\perp
}^{\prime }-q_{\parallel }^{2}\tilde{P}_{\parallel }^{\prime }\right) \psi
^{2}$ and $P_{2}=-\left[ \tilde{P}_{\parallel 0}^{\prime }\left( \varpi
-1\right) +2\varpi \psi _{\parallel 0}^{\prime }\right] $ with $\tilde{P}%
_{\parallel 0}^{\prime }=\left. \mathrm{d}\tilde{P}/\mathrm{d}q_{\parallel
}^{2}\right\vert _{q=0}$.

In the following section, in the same manner as in \cite{5744}, I look for
the fixed point solutions of eqs (\ref{eq:1},\ref{eq:2}) ($\mathrm{\dot{v}}%
_{1}=\mathrm{\dot{v}}_{2}=0$) which satisfy the following regular behavior
at large $\bar{\phi}$:

\begin{eqnarray}
\mathrm{v}_{1\mathrm{asy}} &=&G_{1}\bar{\phi}^{\theta _{1}}+\theta
_{1}G_{1}^{2}\bar{\phi}^{2\theta _{1}-1}+\cdots \\
\mathrm{v}_{2\mathrm{asy}} &=&-\psi _{\parallel 0}^{\prime }\theta _{1}G_{1}%
\bar{\phi}^{\theta _{1}-1}+\cdots +G_{2}\bar{\phi}^{\theta _{2}}+\cdots
\end{eqnarray}%
in which $G_{1}$ and $G_{2}$ are arbitrary constants, $\theta _{1}=\frac{%
d_{m}-2\varpi }{d_{m}+2\varpi }$ and $\theta _{2}=-4\frac{\theta +\varpi }{%
d_{m}+2\varpi }$.

I am interested also in the study of the equations linearized in the
vicinity of a fixed point $\mathrm{v}_{i}^{*}$ (eigenvalue equations). By
setting $\mathrm{v}_{i}=\mathrm{v}_{i}^{*}+\varepsilon \mathrm{e}^{\lambda t}%
\mathrm{g}_{i}$ (with $t=-\ln \ell $) and retaining the linear term in $%
\varepsilon $, the eigenvalue equations read: 
\begin{eqnarray}
\mathrm{g}_{1}^{\prime \prime } &=&\left( \lambda -d_{\phi }+2\varpi \mathrm{%
v}_{1}^{*\prime }\right) \mathrm{g}_{1}+\left( \tilde{d}_{\phi }\bar{\phi}%
+2\varpi \mathrm{v}_{1}^{*}\right) \mathrm{g}_{1}^{\prime }-P_{1}\mathrm{g}%
_{2}^{\prime }  \label{eq:VP1} \\
\mathrm{g}_{2}^{\prime \prime } &=&\left[ \lambda +2\left( \varpi +\theta
+2\varpi \mathrm{v}_{1}^{*\prime }\right) \right] \mathrm{g}_{2}+\left( 
\tilde{d}_{\phi }\bar{\phi}+2\varpi \mathrm{v}_{1}^{*}\right) \mathrm{g}%
_{2}^{\prime }  \notag \\
&&+2\varpi \mathrm{v}_{2}^{*\prime }\mathrm{g}_{1}+2\left( 2\varpi \mathrm{v}%
_{2}^{*}-P_{2}\mathrm{v}_{1}^{*\prime }+\theta \psi _{\parallel 0}^{\prime
}\right) \mathrm{g}_{1}^{\prime }  \label{eq:VP2}
\end{eqnarray}
the interesting solutions of which must satisfy the following large $\bar{%
\phi}$ behavior (with $\varkappa _{1}=\frac{d_{m}-2\varpi -2\lambda }{%
d_{m}+2\varpi }$, $\varkappa _{2}=-2\frac{2\theta +2\varpi +\lambda }{%
d_{m}+2\varpi }$): 
\begin{eqnarray}
\mathrm{g}_{1asy} &=&S_{1}\bar{\phi}^{\varkappa _{1}}+\cdots \\
\mathrm{g}_{2asy} &=&-\psi _{\parallel 0}^{\prime }S_{1}\varkappa _{1}\bar{%
\phi}^{\varkappa _{1}-1}+S_{2}\bar{\phi}^{\varkappa _{2}}+\cdots
\end{eqnarray}

Notice that, since $\ell $ is chosen to be the scaling parameter in the $%
\perp $ direction ($\ell =\ell _{\perp }$) then the eigenvalues $\lambda $
stand for those associated with the $\perp $ direction. Specifically, the
highest positive eigenvalue will be associated to $1/\nu _{L2}$.

\section{Numerical study}

\subsection{Preliminary}

Before looking specifically at the numerical study of (\ref{eq:1},\ref{eq:2}%
) for $m=1$ (uniaxial Lifshitz systems), it is worthwhile to make some
general remarks and comments:

\begin{enumerate}
\item Eqs (\ref{eq:1},\ref{eq:2}) differ from those of the ordinary critical
point up to $O\left( \partial ^{2}\right) $ (see \cite{5744}) essentially
for $\theta \neq 1$ and $d$ replaced by $d_{m}$ [eq (\ref{eq:dm})]. Since $%
\theta =1/2$ in the order presently considered they are essentially the same
if one considers $d_{m}$ as an effective value of $d$. This remark is
important since one must find a fixed point with one supplementary direction
of instability compared to the ordinary critical point [the Lifshitz point
is a multicritical ---tricritical indeed--- point, see eq (\ref%
{eq:generalscaling})] starting with practically the same equations! I
explain below how this is possible.\label{Rem1}

\item The classical dimensional analysis of the Lifshitz system is identical
to that of the ordinary system provided that one changes $d$ into $d_{m}$
with $\theta =1/2$. So, for the Lifshitz criticality, the ordinary upper
critical dimension $d^{\left( u\right) }=4$ becomes $d_{m}^{\left( u\right)
}=4$ that is to say $d_{L}^{\left( u\right) }=4+\frac{m}{2}$ (hence the $%
\epsilon _{L}$-expansion with $\epsilon _{L}=4-d+\frac{m}{2}$); and, for the
Lifshitz tricriticality, the ordinary upper critical dimension $d_{\mathrm{%
tri}}^{\left( u\right) }=3$ becomes $d_{m}^{\left( u\right) }=3+\frac{m}{2}$
(hence the $\epsilon _{Lt}$-expansion with $\epsilon _{Lt}=3-d+\frac{m}{2}$).

\item Below I study the uniaxial Lifshitz system ($m=1$). Since at the
leading order of the derivative expansion one has $\theta =1/2$, then the
effective value of $d$ (in the sense of remark \ref{Rem1}) is $d_{m}=2.5$.
One knows that in the ordinary case for $d<3$ several fixed points can
co-exist. One thus expects the same in the present study allowing for the
presence of several fixed points controlling different critical points.

\item The Lifshitz critical and \emph{tricritical} points may be studied
from equations (\ref{eq:1},\ref{eq:2}), only the fixed point considered is
different in the two cases (it is a question of number of unstable
directions which differentiates the two). Then one understands well why
Nicoll et al \cite{5683} had not the right number of couplings in their $%
\epsilon $-expansion: they had considered only one equation [similar to eq (%
\ref{eq:1}) with $\mathrm{v}_{2}$ lacking]. In fact as stressed in \cite%
{5680} they had forgotten the coupling associated with the operator $\left(
\partial _{\parallel }\phi \right) ^{2}\phi ^{2}$. In the present study such
a term is included within the function $\mathrm{v}_{2}$ (i. e. $Z_{\parallel
}\left( \phi \right) $) and eq (\ref{eq:2}). Notice that $Z_{\parallel
}\left( \phi \right) $ induces supplementary terms in the action and
especially $\left( \partial _{\parallel }\phi \right) ^{2}\phi ^{4}$ which
has not been considered even in \cite{5680} while it is associated with a
marginal coupling at the physical dimension $d=3$ ($d_{m}=2.5$).
\end{enumerate}

In order to numerically study eqs (\ref{eq:1},\ref{eq:2}) I must choose the
cutoff function $\tilde{P}$ and the redundant function $\psi $. But, at this
low order of approximation, the precise forms of these functions have no
great signification. In fact I have explicitly considered (similarly to \cite%
{5744}): 
\begin{eqnarray}
\tilde{P}\left( q_{\parallel }^{2},q_{\parallel }^{2}\right) &=&\mathrm{e}%
^{-a\left( q_{\parallel }^{4}+q_{\perp }^{2}\right) } \\
\psi \left( q^{2}\right) &=&\frac{1}{1+bq^{2}}
\end{eqnarray}
with $a$ and $b$ two parameters. The fourth power of $q_{\parallel }$ in the
cutoff function has been imposed after having observed that no fixed point
exists if $\tilde{P}_{\parallel }^{\prime }\sim q_{\parallel }^{2}$ at small 
$q_{\parallel }$. With this cutoff function, it appears that the results are
insensitive to the value of $a$. As for $b$, its actual significance is to
play with the breakdown of the reparametrization invariance introduced by
the approximation, since at this low order the invariance cannot be studied
(because both $Z_{\perp }$ and $A$ are fixed), I set $b=0$ in the following.
The remaining of the numerical study follows the same lines as in \cite{5744}%
: I consider the shooting method starting from a value $\bar{\phi}_{0}$ at
which point the large $\bar{\phi}$ behavior of the regular solutions of eqs (%
\ref{eq:1},\ref{eq:2}) is imposed then I try, with the Newton-Raphson
algorithm, to find solutions that reach the origin $\bar{\phi}=0$ with some
conditions satisfied. For the fixed point, the conditions are conditions of $%
\mathrm{Z}_{2}$-symmetry: 
\begin{eqnarray}
\mathrm{v}_{1}(0) &=&0  \label{eq:FPcond1} \\
\mathrm{v}_{2}^{\prime }(0) &=&0  \label{eq:FPcond2}
\end{eqnarray}

I find at least two fixed points of interest (other fixed points coexist but
they are less stable): one corresponds to $G_{1}=-2.385$ and the second one
to $G_{1}=-5.031$. For all the fixed points, $G_{2}=0$ and $\mathrm{v}_{2}$
vanishes exactly.

The first fixed point is like the customary fixed point that one encounters
in studying the ordinary critical point and which is called the
Wilson-Fisher fixed point \cite{439}. However, presently it controls the
Lifshitz critical point and possesses one direction of instability more that
the Wilson-Fisher fixed point (see below). This is possible since eqs (\ref%
{eq:1},\ref{eq:2}) are like the $O\left( \partial ^{2}\right) $-equations of
the ordinary case in which $\eta $ would be arbitrarily fixed to zero. In
that case the reparametrization invariance is strongly violated and the
zero-eigenvalue usually associated with the (satisfied) invariance becomes a
positive eigenvalue.

The second fixed point is candidate for representing the Lifshitz
tricritical fixed point, the study of the eigenvalue equations will give the
answer.

\subsection{Study of the eigenvalue equations}

\subsubsection{First fixed point and the Lifshitz critical point}

I look for the eigenvalue equations (\ref{eq:VP1},\ref{eq:VP2}) for the
first fixed point and with the $\mathrm{Z}_{2}$-symmetry imposed (i. e. $%
\mathrm{g}_{1}(0)=0$ and $\mathrm{g}_{2}^{\prime }(0)=0$) and I find two
positive eigenvalues: 
\begin{equation}
\lambda _{1}=1.5785,\qquad \lambda _{2}=1.1557
\end{equation}
yielding: 
\begin{equation}
\nu _{L2} =\frac{1}{\lambda _{1}}=0.6335,\qquad \varphi =\frac{\lambda _{2}}{%
\lambda _{1}}=0.7322
\end{equation}

The first negative eigenvalue is found to be: 
\begin{equation}
\lambda _{3}=-0.8916
\end{equation}
giving the correction exponent $\omega _{L2}$: 
\begin{equation}
\omega _{L2}=-\lambda _{3}=0.8916
\end{equation}

Let me remind you the best results at order $O\left( \epsilon ^{2}\right) $ 
\cite{4708} once $\epsilon =3/2$ is set in the expansions: $\nu _{L2}=0.625$%
, $\varphi =0.625$, $\omega _{L2}=1.5$ at $O\left( \epsilon \right) $; $\nu
_{L2}=0.709$, $\varphi =0.677$, $\omega _{L2}=0.414$ at $O\left( \epsilon
^{2}\right) $ (see the other kinds of estimates in \cite{4708}). One sees
that, regarding the very low order of the approximation, the present results
are very good. But one advantage of the ERGE is that one obtains easily the
eigenvalues in the $\mathrm{Z}_{2}$-asymmetric case, it suffices to impose
the conditions $\mathrm{g}_{1}^{\prime }(0)=0$ and $\mathrm{g}_{2}(0)=0$ and
one finds: 
\begin{eqnarray}
\breve{\lambda}_{1} &=&2.25=\frac{d_{m}+2-\eta _{L2}}{2} \\
\breve{\lambda}_{2} &=&0.25=\frac{d_{m}-2+\eta _{l2}}{2} \\
\breve{\lambda}_{3} &=&-0.5821=-\breve{\omega}_{L2}
\end{eqnarray}

One observes that, since $\breve{\omega}_{L2}<\omega _{L2}$ the corrections
to scaling are, in fact, dominated by the asymmetry. This is a new result.

\subsubsection{Second fixed point and the supposed Lifshitz tricritical point%
}

In the same way as above, in the symmetric case I find the following
eigenvalues: 
\begin{eqnarray}
\lambda _{1} &=&1.984398,\qquad \lambda _{2}=1.050316,\qquad \lambda
_{3}=0.885100,  \notag \\
\lambda _{4} &=&0.068527,\qquad \lambda _{5}=-1.352355  \label{eq:intruder}
\end{eqnarray}%
with four positive answers instead of only three if the fixed point had to
control a Lifshitz tricritical point as expected (the other fixed points
have even more positive eigenvalues). The $O\left( \epsilon \right) $
estimates gave for the three positive eigenvalues \cite{5680}: $\lambda
_{1}=1.86$--$1.87$, $\lambda _{2}=0.857$--$0.875$ and $\lambda _{3}=0.643$--$%
0.737$ according to whether one resums the series or its inverse. The
highest negative eigenvalue was estimated to be $\lambda _{y_{1}}=-1$ \cite%
{5674}. Hence the intruder eigenvalue in (\ref{eq:intruder}) is $\lambda
_{4} $ the closest to zero. A simple dimensional analysis allows to
understand how it can appear and why it has not been accounted for in the $%
\epsilon $-expansion framework.

The two terms $c_{1}\int \mathrm{d}^{d}x\phi ^{6}$ and $c_{2}\int \mathrm{d}%
^{d}x\left( \partial _{\parallel }\phi \right) ^{2}\phi ^{2},$ mentioned at
the beginning of section \ref{section I} as contributing to the action
relevant to the study of the Lifshitz tricritical point in the framework of
the $\epsilon $-expansion \cite{5680,5685}, have couplings of respective
classical dimension $\left[ c_{1}\right] =2\left( 3-d_{m}\right) $ and $%
\left[ c_{2}\right] =3-d_{m}$ and thus can (must) be considered together has
small quantities with $\epsilon _{Lt}=3-d_{m}$. They indeed destabilize the
gaussian fixed point and they are at the origin of a new nontrivial fixed
point of order $\epsilon _{Lt}$ in the same way as the $\phi ^{4}$ coupling $%
u$ destabilizes the gaussian fixed point when $d=4-\epsilon $. A difference
is that it is a tricritical point with the effect of $u$ inhibited. In the
ordinary case, \emph{only} the $\phi ^{6}$ coupling has to be considered,
moreover, since it has the dimension $3-d$, it is marginal at the physical
dimension (the coupling parameter is small) and the study may be done
perturbatively: the gaussian fixed point is actually not destabilized and
the effective exponents are classical up to logarithms. In the present case
of the Lifshitz tricritical point, the physical dimension is not $d_{m}=3$
but instead $d_{m}=2.5$ and the fixed point of interest is no longer the
Gaussian fixed point but a \emph{nontrivial} fixed point (the second fixed
point of the present study). If only the two couplings $c_{1}$ and $c_{2}$
had to be considered ---as effectively supposed in \cite{5680}--- then one
could still use the perturbative approach as in the well known case of the
Wilson-Fisher fixed point \cite{439} relevant to the study of the ordinary
critical point. Unfortunately, at the physical dimension $d_{m}=2.5$ a third
term appears to have a marginal coupling, it is $c_{3}\int \mathrm{d}%
^{d}x\left( \partial _{\parallel }\phi \right) ^{2}\phi ^{4}$ with $\left[
c_{3}\right] =5-2d_{m}$. One does not know whether it is marginally relevant
or marginally irrelevant but one thing is certain: the answer cannot be done
perturbatively because at the dimension where its effect is small ($%
d_{m}=2.5 $), the fixed point of interest (created by the effects of $c_{1}$
and $c_{2} $) is no longer close to the Gaussian fixed point since this
would imply $d_{m}\simeq 3$. A nonperturbative study is obliged in that case
and the present one indicates that $c_{3}$ is marginally relevant at the
fixed point of interest which is thus unstable. Consequently, if this result
is maintained at higher orders of the derivative expansion, the so-called
Lifshitz tricritical point is a point of higher criticality. However, since $%
\lambda _{4}$ is small, the observation of almost universal Lifshitz
tricritical behavior is possible.

\textbf{Acknowledgements}: I thank M. Shpot for having encouraged me to look
at Lifshitz critical points and C. Bagnuls for many discussions on this
subject.


\begin{thebibliography}{99}
\bibitem{note1} I would like to draw reader's attention on the fact that if
one performs the changes $V\rightarrow -H$, $\dot{V}\rightarrow -\theta
\partial H/\partial l$, and $\theta =1/\sigma $ for $m=d$ in eq (\ref%
{eq:LPAOrdin}) then one should obtain (for example) eq (1.2) of \cite{4386}
which was intended to generalize the local potential approximation of the
Wilson equation to critical systems with ``propagator exponent'' $\sigma $.
One may easily check that the two equations differ indeed by the factor $%
\varpi $ appearing in front of the ``coarse-graining'' term $\left(
V^{\prime 2}-V^{\prime \prime }\right) $ in (\ref{eq:LPAOrdin}). This is a
direct consequence of the omission of re-normalization factor $\zeta $ of
the field as stressed in \cite{5744}.
\end{thebibliography}
\end{document}